# Derivation of global vegetation biophysical parameters from EUMETSAT Polar System

Francisco Javier García-Haro[a*], Manuel Campos-Taberner[a], Jordi Muñoz-Marí[b], Valero Laparra[b], Fernando Camacho[c], Jorge Sanchez-Zapero[c] and Gustau Camps-Valls[b]

[a]Earth Physics and Thermodynamics Department, Faculty of Physics, Universitat de València, Dr. Moliner,
46100 Burjassot, València (Spain).

[b]Image Processing Laboratory (IPL), Universitat de València, C/ Catedrático José Beltrán, 2, 46980 Paterna, València (Spain).

[c]Earth Observation Laboratory (EOLAB), Parc Científic de la Universitat de València, Catedrático A. Escardino, 46980 Paterna, València (Spain).

[*]E-mail: j.garcia.haro@uv.es; Tel. +34 963543111

**Abstract**

This paper presents the algorithm developed in LSA-SAF (Satellite Application Facility for Land Surface Analysis) for the derivation of global vegetation parameters from the AVHRR (Advanced Very High Resolution Radiometer) sensor on board MetOp (Meteorological–Operational) satellites forming the EUMETSAT (European Organization for the Exploitation of Meteorological Satellites) Polar System (EPS). The suite of LSA-SAF EPS vegetation products includes the leaf area index (LAI), the fractional vegetation cover (FVC), and the fraction of absorbed photosynthetically active radiation (FAPAR). LAI, FAPAR, and FVC characterize the structure and the functioning of vegetation and are key parameters for a wide range of land–biosphere applications. The algorithm is based on a hybrid approach that blends the generalization capabilities offered by physical radiative transfer models with the accuracy and computational efficiency of machine learning methods. One major feature is the implementation of multi-output retrieval methods able to jointly and more consistently estimate all the biophysical parameters at the same time. We propose a multi-output Gaussian process regression ($GPR_{multi}$), which outperforms other considered methods over PROSAIL (coupling of PROSPECT and SAIL (Scattering by Arbitrary Inclined Leaves) radiative transfer models) EPS simulations. The global EPS products include uncertainty estimates taking into account the uncertainty captured by the retrieval method and input errors propagation. A sensitivity analysis is performed to assess several sources of uncertainties in retrievals and maximize the positive impact of modeling the noise in training simulations. The paper discusses initial validation studies and provides details about the characteristics and overall quality of the products, which can be of interest to assist the successful use of the data by a broad user's community. The consistent generation and distribution of the EPS vegetation products will constitute a valuable tool for monitoring of earth surface dynamic processes.

# 1 Introduction

The Satellite Application Facility for Land Surface Analysis (LSA-SAF) (https://landsaf.ipma.pt) is a dedicated processing center serving the European Organization for the Exploitation of Meteorological Satellites (EUMETSAT). The main purpose of the LSA-SAF is to develop and implement algorithms that allow an operational use of land surface variables taking benefit of remote sensing data from sensors onboard EUMETSAT satellites in order to estimate land surface properties. Although LSA-SAF has been particularly targeted to the meteorological community needs, especially Numerical Weather Prediction, the LSA-SAF products address a much broader community (Trigo *et al*., 2011). Since the end of 2008, the LSA-SAF produces and disseminates a suite of vegetation parameters from SEVIRI/MSG (Spinning Enhanced Visible and Infrared Imager/Meteosat Second Generation) data over the Meteosat disk at two different time resolutions: daily and 10-day: Leaf Area Index (LAI), Fraction of Absorbed Photosynthetically Active Radiation (FAPAR), and Fractional Vegetation Cover (FVC). LAI is a quantitative measure of the amount of live green leaf material present in the canopy. It is defined as half the total area of green elements per unit horizontal ground area (Chen & Black, 1992) accounting for the amount of green vegetation that absorbs or scatters solar radiation. FAPAR accounts for the active radiation absorbed by the canopy in the range of 400-700 nm, and therefore constitutes an indicator of the health and thereby productivity of vegetation (Asner *et al*., 1998). FAPAR depends on the angular position of the Sun and is suitable to quantify $CO_2$ uptake by plants and the water release through evapotranspiration (Martinez *et al*., 2018). LAI and FAPAR have been selected as Essential Climate Variables (ECVs) by the Global Climate Observing System (GCOS) and are key for sustainable climate observations (GCOS, 2011). FVC represents the green vegetation fraction that covers a unit area of horizontal soil, corresponding to the gap fraction seen from the nadir (Bonham, 2013). Unlike FAPAR, FVC does not depend on variables such as the geometry of illumination being thus a good alternative to vegetation indices for monitoring Earth's green vegetation. LAI, FAPAR, and FVC characterize the structure and the functioning of vegetation and are key inputs for a broad variety of biosphere and land applications, from climate, forestry and agriculture to environmental and natural hazards management. These variables play a key role in crop modelling with respect to the simulation of processes, such as photosynthesis, respiration, evapotranspiration and rain interception.

This paper describes the methodology applied for the retrieval of global LAI, FAPAR and FVC from the Advanced Very High Resolution Radiometer (AVHRR) sensor onboard the MetOp (Meteorological–Operational) satellite constellation also known as EUMETSAT Polar System (EPS). Unlike the approach adopted to produce SEVIRI/MSG vegetation products (García-Haro *et al.*, 2016), the proposed algorithm relies on the inversion of Radiative Transfer Models (RTMs) in order to ensure the consistency among derived vegetation parameters. RTMs model the interplay among the radiation, the canopy and the soil, and rely on their physical knowledge capturing the physics of the involved radiation interaction processes. A wide range of radiative transfer models have been proposed in the literature since last four decades (Verhoef, 1984; Verhoef & Bach, 2007). In particular, PROSAIL, which arises from the PROSPECT (Jacquemoud & Baret, 1990) and the SAIL (Verhoef, 1984) coupling, has been successfully used in research studies and implemented in operational bio-physical parameter retrieval chains (Duan *et al.*, 2014; Campos-Taberner *et al.*, 2016; 2017).

Retrieval methods inverting RTMs have been developed to generate operational biophysical products from Earth observation data: CYCLOPES (Carbon cYcle and Change in Land Observational Products from an Ensemble of Satellites) (Baret *et al.*, 2007) products are generated from VEGETATION data inverting the PROSAIL model, the MODIS (MODerate resolution Imaging Spectroradiometer) LAI/FAPAR products are derived from a 3-D radiative transfer model (Myneni *et al.*, 2002; Knyazikhin *et al.*, 1998) defined for eight biomes, while the Copernicus GEOland2 (GEOV1) products (Baret *et al.*, 2013) are derived from SPOT/VEGETATION (*Satellite Pour l'Observation de la Terre*/VEGETATION) & PROBA-V (Project for On-Board Autonomy-V) by fusing and scaling MODIS and CYCLOPES products.

The choice of the most suitable EPS algorithm needs to take into account the reliability of parameters and respective uncertainty estimates, based on the expected accuracy, robustness and timeliness for operational production. The above criteria favor the use of hybrid approaches, which are computationally efficient respect the basic physical rules encoded in RTMs, and are generally free of the convergence problems of physical inversion retrieval methods (Verrelst *et al.*, 2012, 2015; Camps-Valls *et al.*, 2016) associated to the non-linearity of the inverse problem in remote sensing (Camps-Valls *et al.*, 2017). Hybrid methods blend the generalization of physical models with the accuracy and efficiency of (non-parametric) machine learning approaches (Verger *et al.*, 2008; Verrelst *et al.*, 2012; Houborg and McCabe, 2018). Among hybrid methods, Neural Networks (NNs) have been implemented in operational processing chains for retrieving global biophysical parameters (Bacour *et al.*, 2006; Baret *et al.*, 2007) inverting the PROSAIL model. Recently, novel kernel-based algorithms have been introduced in classification and regression problems in remote sensing (Pérez-Suay *et al.*, 2017). Support Vector Regression (SVR) was used for LAI, FVC and evapotranspiration retrieval (Yang *et al.*, 2006; Durbha *et al.*, 2007). Gaussian Processes regression (GPR) was also used for LAI retrieval, outperforming other methods (Lázaro-Gredilla *et al.*, 2014; Campos-Taberner *et al.*, 2016). Moreover, GPR is able to deal with different data and noise, also providing associated confidence intervals for the predictions. Nevertheless, GPR has not been used yet in operational retrieval chains for vegetation parameter estimation at global scale.

Common techniques for retrieving biophysical parameters use single-output methods, which optimize each parameter separately. For example, the CYCLOPES and GEOV1 algorithms use three networks in parallel, one for each biophysical parameter. When parameters are related each other, a single multi-output model is more computationally efficient, and more importantly it should provide improved results in terms of consistency and accuracy. One of the main features of this paper has been the implementation of multi-output methods in which the retrieved biophysical variables share the same model's parameters. Previous works have found that the optimization of the model parameters for the joint retrieval of biophysical parameters leads to better results (Tuia *et al.*, 2011). The present study has also addressed the modelling of noise in RTM simulations, which may reduce the retrieval errors in cases where simulations depart slightly from observations (Baret *et al.*, 2007; Verger *et al.* 2011).

This paper describes the algorithm currently integrated into the LSA-SAF operational system to produce biophysical products from AVHRR/MetOp data using a novel GPR multi-output algorithm (GPRmulti) within a hybrid retrieval approach. There are a number of advantages for using these products: (i) enhanced consistency of the EPS LAI, FAPAR and FVC estimates based on a joint retrieval method, (ii) provision of well-characterized uncertainty on a pixel-by-pixel basis, taking into account the uncertainty of retrieval method, input errors propagation as well as the retrieval conditions, (iii) dissemination as global files in near real time and on a 10-day basis through the project website and EUMETCast, (iv) good completeness and low ratio of missing values in the tropical, subtropical and warm temperate regions, (v) continuity of the processing chain for long-term monitoring, through the adaptation for the second generation of EUMETSAT polar satellites (EPS-SG) and the generation of Climate Data Records (CDR), (vi) a comprehensive documentation regularly updated including quality control and routinely validation studies, (vii) user-oriented activities and feedback by the LSA-SAF user support team, (viii) availability of a set of AVHRR/Metop parameters related with land surface temperature and albedo. Section 2 outlines the main components of the proposed multi-output retrieval chain, describing the GPRmulti principles. Section 3 assesses its performance with regard to other multi-output nonlinear regression versions of NN and KRR (kernel ridge regression) while section 4 assesses the uncertainty of products and investigates the optimal amount of noise in simulated reflectances to enhance the quality of retrievals. Section 5 describes the global EPS products along with useful details about its overall quality. Section 6 discusses initial validation studies. Eventually, section 7 highlights the main conclusions and future developments.

## 2 Algorithm description

The general procedure for deriving biophysical parameters using hybrid methods is to run the RT model first (in direct mode) to build a database of reflectance and associated biophysical parameters representing a broad set of canopy parameterizations. The generated simulations are then used to train a (non-linear) non-parametric regression model through machine learning approaches. The main goal of the proposed algorithm is the inversion of the PROSAIL RTM with a family of proposed multi-output kernel-based retrieval methods and neural networks. The best method in terms of stability, accuracy, and robustness was then implemented into the operational chain for the joint retrieval LAI, FVC and FAPAR maps globally from corresponding EPS surface reflectance data. A general outline of the methodology is shown in Figure 1, and its main ingredients including the PROSAIL RTM, the statistical regression algorithms for model inversion and the inputs are described in the next sections.

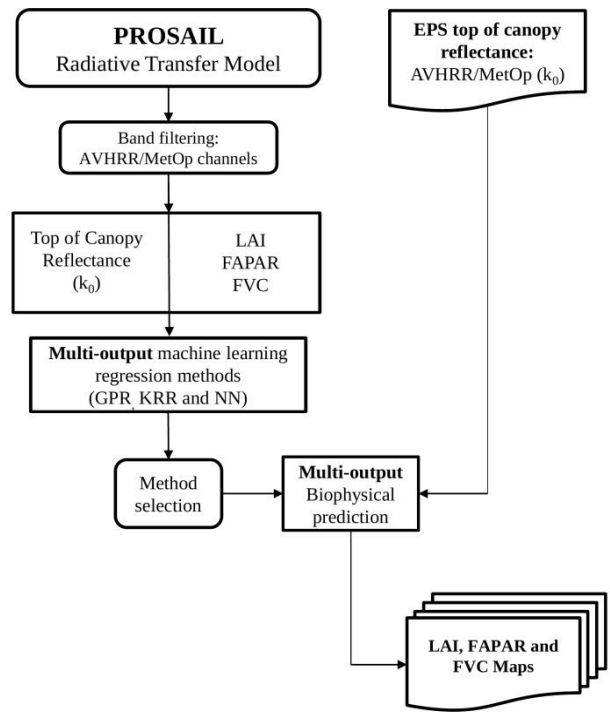

Figure 1. Workflow of the proposed methodology for the joint derivation of the EPS (LAI, FVC and FAPAR) vegetation products.

## 2.1 EUMETSAT AVHRR/MetOp reflectances

EUMETSAT has the operational responsibility for the Europe's first polar orbiting operational meteorological constellation forming the EPS, the first of which (MetOp-A) was successfully launched in 2006, the second (MetOp-B) in 2012, whereas the launch of the third (MetOp-C) is foreseen for 2018. MetOp carries onboard a wide range of sensors, and among them, the AVHRR instrument is the main sensor in charge of providing observations useful for most of the parameters that LSA-SAF supplies.

The AVHRR has a swath of about 2400 km, providing Earth observations with view zenith angles up to about 60º. This sensor offers the capability to observe the whole globe every day at 1.1 km spatial resolution (at nadir) on 6 channels of the electromagnetic spectrum. The algorithm of EPS vegetation products uses as input atmospherically corrected cloud-cleared BRDF (Bidirectional Reflectance Distribution Function) data which is a LSA SAF internal product derived from the albedo algorithm (Geiger *et al*. 2016). Surface reflectance is characterized by the BRDF which describes the appearance of a land surface by its interaction with radiation at a surface point. The algorithm to estimate the BRDF (Geiger *et al*., 2008) applies a semi-empirical reflectance model in order to invert top-of-canopy (TOC) reflectance factor values into a number of parameters ($k_0$, $k_1$, $k_2$) which are associated to dominant light scattering processes (Roujean *et al*. 1992).

The LSA SAF algorithm to retrieve FVC, LAI and FAPAR relies on the normalized spectral reflectance factor, i.e. BRDF $k_0$ parameter, in three EPS channels, centered at about 0.63 μm (red, C1), 0.87 (NIR, C2) and 1.61 μm (MIR, C3) (see

Figure 2). Physically the BRDF $k_0$ parameter corresponds to isotropic reflectance, i.e. reflectance factor values directionally normalized to reference illumination and observation zenith angles of 0°. Retrieval of vegetation parameters requires identification of cloud-free pixels and correction of atmospheric effects. The albedo chain performs the SMAC (Simplified Model for Atmospheric Correction) atmospheric correction (Rahman & Dedieu, 1994) using auxiliary information from the cloud-mask product (CMa) developed by the Nowcasting-SAF (NWC SAF; http://nwcsaf.inm.es/), surface pressure and atmospheric constituents (vapor, ozone) from the ECMWF (European Centre for Medium-Range Weather Forecasts), and aerosol climatology from the Copernicus Atmosphere Monitoring Service (CAMS). Detection of snow in the input albedo observations relies also on the CMa-NWC.

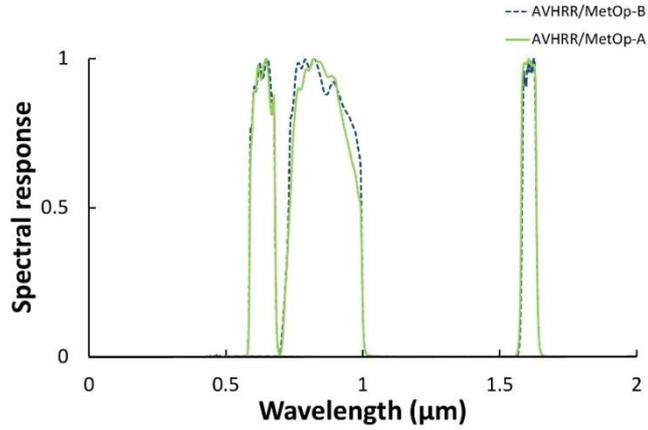

Figure 2. Spectral response functions of AVHRR optical channels (1, 2 and 3), centered at 0.630, 0.865 and 1.610 μm, respectively, onboard MetOp-A and MetOp-B.

## 2.2 PROSAIL simulations

RTMs describe the interaction between radiation and canopy. This allows the simulation of canopy spectral signatures from the leaf, canopy, as well as the background spectral characteristics. In this study, the PROSPECT-5B (Feret *et al*., 2008) and the SAILH (Verhoef, *et al*., 2007) were used for PROSAIL RTM coupling. PROSPECT-5B simulates the directional–hemispherical reflectance of the leaf and its transmittance in the spectral range of 400-2500 nm at 1 nm step with updated absorption coefficients of leaf constituents and leaf refractive index. PROSPECT-5B deals with chlorophylls and carotenoids separately while PROSPECT-4 uniformly treats all photosynthetic pigments. Leaf optical properties are expressed as a function of the mesophyll parameter N (unitless), leaf chlorophyll $C_{ab}$ and carotenoid $C_{ar}$ contents expressed in μg·cm$^{-2}$ as well as the dry matter $C_m$ (g·cm$^{-2}$), relative water $C_{REL}$ (unitless), and leaf brown pigment $C_{bp}$ contents (unitless). SAILH uses as input canopy level parameters such as LAI and average leaf angle (ALA) also incorporating the foliage hot-spot effect (Kuusk, 1985).

The underlying bareground was characterized by a representative set of background spectra (assumed to be Lambertian) multiplied by a brightness parameter ($β_s$) (Baret *et al*., 2007; Claverie *et al*., 2013). Bare areas were selected taking into account the high variability in global soil background reflectance conditions for land surfaces. An appropriate bareground characterization in the PROSAIL simulations is necessary to produce realistic estimates in incomplete canopies, such as sparse canopies or crops during first growing stages (Campos-Taberner *et al*., 2016). Bareground spectra were extracted from an EPS devegetated composited image to reduce the presence of vegetative cover. The composited image was built on an annual time-series by considering all optimal quality reflectance (BRDF $k_0$) and retaining the $k_0$ corresponding to the lowest NDVI value. Bare soils were identified in the border of the convex hull of the EPS channels (green circles in Figure 6). The samples were chosen to be homogeneous based on GLC2000 classification and included mainly bare areas, sparsely vegetated and open shrublands (classes 19, 14 and 12). One major drawback is that GLC2000 was based primarily on SPOT VEGETATION data collected during 2000, and land cover changes could be expected since then. Nevertheless, this dataset was chosen due to its relatively high accuracy, appropriate legend and accurate identification of bare areas (Kaptué-Tchuenté *et al.*, 2011; Pérez-Hoyos *et al.*, 2012). In order to improve the representativeness of the background spectra selected, additional samples were selected in sites identified as bare areas either presenting positive LAI/FAPAR retrievals, or large associated uncertainty. Some identified pixels were further examined based on Landsat 8 imagery. Some samples were also added based on ancillary data and exploiting the availability homogeneous BELMANIP (Benchmark Land Multisite Analysis and Intercomparison of Products) sites (Baret *et al.*, 2006), the LSA-SAF validation stations in Africa and Google Earth imagery (Yu and Gong, 2011). Samples were further verified using the outcomes of the retrieval chain to filter out possible outliers, i.e. LAI > 0.3 and significant seasonality in LAI.

The geometry of the system was characterized by the solar and view zenith angles ($\theta_s$ and $\theta_v$, respectively), and the relative azimuth angle ($\Delta\Phi$), which in our case corresponded to illumination and observation zenith angles of 0°.

For FAPAR, the PROSAIL provides both direct (black sky) and diffuse (white sky) FAPAR components, and parameterizes the diffuse to direct fraction (*skyl*) based on the François *et al.*, (2002) formulation. Since instantaneous FAPAR depends on sun position, daily-integrated FAPAR simulations were computed by integrating FAPAR over the day (i.e. from sunrise to sunset). This daily-integrated green FAPAR is demanded for users since the majority of the primary productivity models which use FAPAR operate at daily time step. FVC of a turbid medium is computed within PROSAIL as $FVC = 1 - t_{oo}$, where $t_{oo} = \exp(-C_0 \times LAI)$, being $C_0$ the extinction coefficient for direct observed flux which depends on the leaf angle distribution from average leaf angle (assuming an ellipsoidal distribution).

Spatial heterogeneity is addressed assuming each pixel is composed by a mixture of pure vegetation (*vCover*) and bare soil (*1-vCover*) fractions. We first simulated "pure vegetation" pixels (i.e. R$_{veg}$) based on PROSAIL using as input the LAI distribution in table 1. After that, in order to account for the mixing effect, the pixel reflectance was expressed as a linear combination, R= R$_{veg}$×*vCover*+R$_{soil}$×(*1-vCover*), where R$_{soil}$ refers to bare soil background. The three EPS parameters (LAI, FVC, and FAPAR) of mixed pixel were multiplied by *vCover* to obtain the biophysical values in the simulations database.

One advantage of the use of simulated databases is that site-specific information about parameter distributions is not essential for an appropriate model parameterization. In fact, detailed statistical information is usually unavailable, particularly in a global context. The PROSAIL parameters were randomly generated according to the statistical distributions of Table 1, which were similar to those adopted in other literature studies (Baret *et al.*, 2007; Claverie *et al.*, 2013). Although prior information on the distributions of the leaf and canopy variables may increase the reliability of solutions because of the ill-posed nature, fairly accurate estimates of canopy variables are possible based on a generic training dataset (Verger *et al.*, 2011). It should be noted that brown pigments were intentionally fitted to zero in other to account only for photosynthetic elements of the canopy.

Table 1. Ranges and distributions of the PROSAIL parameters adopted in the EPS retrieval chain. (*) In the case of broadleaved forests (based on GLC2000 classification), the mean value was increased to 4.5. (**) A 5% of spectra representative of pure background (vCover=0) were included to account for bare areas.

|        | Parameter | Min | Max | Mean | Std | Type |
|--------|-----------|-----|-----|------|-----|------|
| Canopy | LAI (m$^2$/m$^2$) | 0 | 8 | 3.5 | 4 | Gaussian(*) |
|        | ALA (°) | 35 | 80 | 62 | 12 | Gaussian |
|        | Hotspot | 0.1 | 0.5 | 0.2 | 0.2 | Gaussian |
|        | vCover | 0.3 | 1 | 0.99 | 0.2 | Truncated gaussian(**) |
| Leaf   | N | 1.2 | 2.2 | 1.5 | 0.3 | Gaussian |
|        | $C_{ab}$ (µg·cm$^{-2}$) | 20 | 90 | 45 | 30 | Gaussian |
|        | $C_{ar}$ (µg·cm$^{-2}$) | 0.6 | 16 | 5 | 7 | Gaussian |
|        | $C_{dm}$ (g·cm$^{-2}$) | 0.005 | 0.03 | 0.015 | 0.008 | Gaussian |
|        | $C_{REL}$ | 0.6 | 0.85 | 0.75 | 0.1 | Gaussian |
|        | $C_{bp}$ | 0 | 0 | 0 | 0 | - |
| Soil   | $\beta_s$ | 0.1 | 1 | 0.8 | 0.6 | Gaussian |

Over dense canopies, reflectance becomes insensitive to LAI variations, since the canopy lower layers are unseen by the satellite sensor, leading to the well-known LAI saturation effect. To better constrain the model inversion and hence the quality of the result over very dense canopies, the mean of LAI distribution was enlarged in Broadleaved forests. This biome is known to present the largest LAI values (Camacho *et al*. 2017a).

The simulating space was more evenly distributed using a Latin hypercube (Mckay *et al*., 2000) sampling. The training database was finally composed of 2950 cases of reflectances in the AVHRR channels and the corresponding biophysical parameters accounting for any combination of the PROSAIL parameters. Finally, top of canopy reflectances were simulated for each wavelength and filtered in accordance with the spectral response of the AVHRR channels (see

Figure 2).

Model inversion performance is sensitive to both satellite reflectance uncertainties and RTM inadequacies. Adding noise in reflectance simulations is a way of accounting for different noise sources in the input such as incomplete cloud screening, atmospheric correction, BRDF normalization and radiometric calibration as well as the suitability of the RTMs to describe photon transport in vegetation mediums (Atzberger *et al*., 2015; Baret *et al*., 2007; Le Maire *et al*., 2008). In the present work, a white Gaussian noise was added to the reflectances of the PROSAIL simulations:

$$R_{train}(\lambda) = R_{sim}(\lambda) + \mathcal{N}(0, \sigma^2(\lambda)), \qquad (1)$$

where $R_{train}(\lambda)$ represent reflectance values for band λ used as input in the retrieval algorithm, as obtained adding to PROSAIL simulations, $R_{sim}(\lambda)$, a normal distribution of noise with standard deviation σ(λ). Since the covariance matrix of normalised directional reflectances $k_0$ is diagonal, with very similar $k_0$ errors (Err($k_0$)) in the three EPS bands, the noise standard deviation was assumed to be wavelength independent. The choice of the optimal amount of noise was driven by a sensitivity assessment (Section 4.3).

## 2.3 Inversion methods

In this paper, we follow a hybrid inversion approach. Essentially, we propose the inversion of RTMs using machine learning methods trained on the generated input-output (reflectance-parameters) data pairs generated by PROSAIL. We evaluated three non-linear regression methods: neural networks (NN), kernel ridge regression (KRR), and Gaussian Process Regression (GPR). For the joint retrieval of LAI, FAPAR, and FVC, we propose multioutput versions for all these methods.

### 2.3.1 *Neural networks (NNs)*

Neural networks rely on the combination of non-linear processing units, called nodes or neurons, into a layered structure which is fully interconnected. The networking is able to model non-linear relations and has been by far the most common approach for decades in many application domains in general and for biophysical parameter retrieval in particular. NN have been used in many hybrid inversion experiments for retrieval of canopy parameters, see Bacour *et al*. (2006) and Baret *et al*. (2013) for some key applications in the field.

Each neuron in a network performs a linear regression followed by a non-linear squashing (sigmoid-like) function (Haykin, 1999). Neurons of different layers are interconnected by scalar weights that are adjusted in the training stage. We used the standard multilayer perceptron trained to minimize the least squares cost using the backprop algorithm. Several network parameters impact the solution so they have to be adjusted: number of hidden layers, and nodes per layer, the learning rate learning algorithm, and eventually some regularization parameters.

### 2.3.2 Kernel ridge regression (KRR)

The Kernel ridge regression (KRR) (Shawe-Taylor & Cristianini, 2004) is the nonlinear (kernel) version of the linear regression model. The KRR solves the least squares linear regression problem with data mapped into a high-dimensional feature space $\mathcal{H}$. The mapping of point (spectral vector) $\mathbf{x}_i$ to that space is represented as $\phi(\mathbf{x}_i)$. When all spectral data are collectively grouped in a matrix, the prediction function is expressed by $\mathbf{Y} = \mathbf{\Phi W}$ and, assuming additive noise, $\hat{\mathbf{Y}} = \mathbf{Y} + \mathbf{E}$, one typically assumes Gaussian noise of zero mean and standard variance $\sigma_n^2$, thus expressing $\mathbf{E} \sim \mathcal{N}(\mathbf{0}, \sigma_n^2 \mathbf{I})$.

The (regularized) squared loss function is $\mathcal{L} = \|\mathbf{\Phi W} - \mathbf{Y}\|^2 + \lambda \|\mathbf{W}\|^2$, which has to be minimized with respect to model weights $\mathbf{W}$. The main problem is that $\mathbf{\Phi}$ is generally not known explicitly. However, the problem is solvable as the solution can be expressed in terms of the dot products of $\mathbf{\Phi}$, and by applying the representer theorem, we only need to operate with similarities implicitly reproduced by a kernel function, $\mathrm{K}(\mathbf{x}_i, \mathbf{x}_j) = \phi(\mathbf{x}_i)\phi(\mathbf{x}_j)^\top$. Similarities between all $n$ input spectra can be grouped in a kernel matrix $\mathbf{K} = \mathbf{\Phi \Phi}^\top$. Therefore, the solution reduces to $\boldsymbol{\alpha} = (\mathbf{K} + \lambda \mathbf{I})^{-1} \mathbf{Y}$, where the new weights $\boldsymbol{\alpha}$ can be used for prediction on new test examples, $\mathbf{X}_*$:

$$\hat{\mathbf{Y}}_* = \mathbf{\Phi}_* \mathbf{W} = \mathbf{\Phi}_* \mathbf{\Phi}^\top \boldsymbol{\alpha} = \mathbf{K}_* \boldsymbol{\alpha} = \mathbf{K}_* (\mathbf{K} + \lambda \mathbf{I})^{-1} \mathbf{Y} \qquad (2)$$

where the matrix $\mathbf{K}_*$ compares test and training samples. In our experiments we used the Gaussian radial basis function (RBF) kernel: $K(\mathbf{x}_i, \mathbf{x}_j) = \exp\left(-\frac{\|\mathbf{x}_i - \mathbf{x}_j\|^2}{2\sigma^2}\right)$. Two parameters need to be optimized: the kernel lengthscale $\sigma$ and the regularization parameter $\lambda$.

### 2.3.3 Gaussian process regression (GPR)

Gaussian processes (GPs) are statistical methods for classification (Kuss & Rasmussen, 2005), regression (Rasmussen & Williams, 2006), and dimensionality reduction (Lawrence, 2005). As we will see, the prediction model for GPR and KRR are essentially the same, but GPs provide a probabilistic approach to kernel regression. A multivariate Gaussian prior is assumed to model some unobserved latent functions, and their respective likelihood. GPR approximates outputs (in our case, the biophysical parameter) as the sum of some unknown latent function of the inputs $f(\mathbf{x})$ (in or case, the normalized reflectance ($k_0$) on the three EPS bands) plus constant Gaussian noise, i.e.

$$y = f(\mathbf{x}) + \varepsilon_n, \ \varepsilon_n \sim \mathcal{N}(0, \sigma_n^2) \tag{3}$$

Now, a zero mean Gaussian process prior is selected as the latent function, that is $(\mathbf{x}) \sim \mathcal{GP}(0, k_\theta)$, where $k_\theta$ is a covariance (kernel) function parametrized by θ. The noise $\varepsilon_n$ is assumed to follow a Gaussian distribution prior, where the noise power is given by the hyperparameter $\sigma_n^2$. Given the GP priors, samples drawn from $f(\mathbf{x})$ at $\mathbf{x}_i = \{x_i^1, x_i^2, x_i^3\}_{i=1}^N$ (in our case, $N$ is the number of training reflectances simulated with PROSAIL) follow a joint multivariate zero mean Gaussian distribution defined by the covariance matrix $\mathbf{K}$, where $[\mathbf{K}]_{ij} = k_\theta(\mathbf{x}_i, \mathbf{x}_j)$.

Given a training dataset $D \equiv \{\mathbf{x}_n, y_n | n = 1, \dots N\}$ and a new input test $\mathbf{x}_*$ its corresponding output $\mathbf{y}_*$ can be obtained given that the GP induces a predictive distribution described by the equations:

$$p(y_*|\mathbf{x}_*, D) = N(y_*|\mu_{GPR*}, \sigma_{GPR*}^2) \tag{4}$$

$$\mu_{GPR*} = \mathbf{k}_*^\top (\mathbf{K} + \sigma_n^2 \mathbf{I})^{-1} \mathbf{y} = \mathbf{k}_*^\top \alpha \tag{5}$$

$$\sigma_{GPR*}^2 = \sigma^2 + k_{**} - \mathbf{k}_*^\top (\mathbf{K} + \sigma_n^2 \mathbf{I})^{-1} \mathbf{k}_*, \tag{6}$$

where $\mathbf{k}_* = [k(x_*, x_1), \dots, k(x_*, x_N)]$ is an $N \times 1$ vector and $k_{**} = k(x_*, x_*)$. The GPR model offers a full posterior probability establishing a relationship between the input and the output variables, from which one can compute pointwise estimations, $\mu_{GPR*}$ and also confidence estimates $\sigma_{GPR*}^2$. The input is, in our case, the spectra on the three EPS $k_0$ bands $\mathbf{x}_i \in \mathbb{R}^3$. If the aim is to retrieve a single parameter, the output is the biophysical parameter of interest (e.g., LAI) $\mathbf{y}$. In this case, given a test spectrum $\mathbf{x}$ the input-output relation is now given by:

$$\hat{y} = f(\mathbf{x}) = \sum_{i=1}^N \alpha_i k_\theta(\mathbf{x}_i, \mathbf{x}) + \alpha_0, \tag{7}$$

where $\alpha_i$ is the weight assigned to each training spectrum $\mathbf{x}_i$, $\alpha_0$ is the bias term, and $k_\theta$ is the kernel (covariance) function evaluating the similarity between $\mathbf{x}$ and all the $N$ training spectra. For the GP model the automatic relevance determination (ARD) kernel was used:

$$K(\mathbf{x}_i, \mathbf{x}_j) = v \exp\left(-\sum_{b=1}^B \frac{(\mathbf{x}_i^b - \mathbf{x}_j^b)^2}{2\sigma_b^2}\right) + \sigma_n^2 \delta_{ij}, \tag{8}$$

where $v$ is a scaling factor, $\sigma_n$ accounts for the noise standard deviation, $B$ is the number of bands (in our case, B=3), and $\sigma_b$ can be related to the relevance (spread) of each spectral band $b$ (low values of $\sigma_b$ indicate a higher impact of band $b$ on the predictive function). Model hyperparameters are collectively grouped in $\boldsymbol{\theta} = [v, \sigma_n, \sigma_1, \dots, \sigma_b]$ and can be estimated by maximizing the marginal log-likelihood (Rasmussen & Williams, 2006):

$$\log p(\mathbf{y}|\mathbf{x}_i, \theta) = -\frac{1}{2}\mathbf{y}^\top(\mathbf{K} + \sigma_n^2 \mathbf{I}_n)^{-1}\mathbf{y} - \frac{1}{2}\log|\mathbf{K} + \sigma_n^2 \mathbf{I}_n| - \frac{N}{2}\log(2\pi) \qquad (9)$$

### 2.3.4 *Joint retrieval of FVC, LAI, and FAPAR: extension to multi-output GPR*

When the goal is to predict multiple variables, the construction of a unique model able to do all the prediction simultaneously may be advantageous, both in computational terms, prediction accuracy and consistency of the predictions. In particular, predicting all three biophysical parameters at the same time could leverage an improved consistency of the estimates since the biophysical parameters are physically linked. In order to achieve this goal, the three single output methods considered above have been adapted to derive jointly the three EPS vegetation parameters (i.e, LAI, FAPAR and FVC), which is one of the main features of the proposed algorithm. This approach was achieved formulating multi-output versions of the NN (NN$_{multi}$), KRR (KRR$_{multi}$), and GPR (GPR$_{multi}$). It should be noted that in the case of NN, the approach is a multioutput algorithm *per se* given its characteristics (i.e. connected layers, weights, and biases). The NN hyperparameters were the number of neurons and hidden layers (to control model complexity we evaluated networks with just one hidden layer and between 2 to 30 hidden neurons). The learning rate was varied in the range 0.001-0.1 in log-scale. Different initializations of the weights were tested to check for consistency of the models.

For the case of kernel methods KRR and GPR, the algorithms can be set up to cope with multioutput regression problems adapting the kernel hyperparameters for a unique kernel which is able to deal with all the outputs. In this paper, the optimization of the hyperparameters for the unique kernel was done either by cross validation or by maximizing the marginal likelihood in the case of KRR$_{multi}$ and GPR$_{multi}$, respectively. For the KRR$_{multi}$ model, it was varied (in log-scale) the regularization parameter $\sigma$ between $10^{-5}$ and $10^{-2}$ and the kernel lengthscale $\lambda$ in the range 0.1-10 times the average distance between all training points. In the standard single-output GPR case we inferred the hyperparameters in $\boldsymbol{\theta} = [v, \sigma, \sigma_1 \ldots \sigma_b]$ and model weights using an optimization of the evidence (Eq. 7), whereas in the GPR$_{multi}$ we need to optimize the parameters taking into account that $\mathbf{y} \in \mathbb{R}^3$ instead $\mathbf{y} \in \mathbb{R}$ where $D$ is the total number of outputs (in our case $D=3$). In order to do so we define a global cost function that sumarizes all the cost functions (one per output) into a global cost function (which will be scalar). Here we propose the global cost function $C$ just as the squared sum of the standard GPR cost funtion of each output:

$$C = \sum_{d=1}^{D} \log p(\mathbf{y}_d|\mathbf{x})^2 = \sum_{d=1}^{D}\left(-\frac{1}{2}\mathbf{y}_d^\top \boldsymbol{\alpha} - \sum_{j=1}^{N}\log L_{jj} - \frac{N}{2}\log(2\pi)\right)^2, \qquad (10)$$

where $L_{jj}$ accounts for Cholesky factorization of the covariance matrix for every output.

## 3 Assessment of the algorithm

The performance of the multi-output inversion methods was evaluated generating 2950 data pairs (reflectances-LAI, FAPAR, and FVC values) with PROSAIL. We used 80% of the simulations for training and we assessed the obtained results in the remaining 20% of the samples (test set unused during model training). One advantage of the multi-output methods for biophysical parameter retrieval with regard to single output approaches is the improved capability in the preservation of the covariance matrix among outputs, which in our case was in line with other multi-output studies using multiple SVR (Tuia *et al*., 2011). In addition, results indicated that the multi-output versions show a slight gain in accuracy with regard to their respective single output versions (see Table 2). The gain in accuracy of every method and biophysical parameter was computed measuring the reduction in root mean square error (RMSE):

$$RMSE\ gain(\%) = \frac{(RMSE_{single-output} - RMSE_{multi-output})}{RMSE_{single-output}} \times 100. \qquad (11)$$

This gain can be related to the fact that, the multi-output optimization links predictions in such a way that the relationships among the biophysical parameters are better described, which helps in regularizing the training procedure obtaining more robust models and therefore improving the accuracy of the estimates.

Table 2. Accuracy improvement of the multi-output methods with regard to their single output versions for LAI, FVC, and FAPAR, respectively.

| Method | LAI RMSE gain (%) | FVC RMSE gain (%) | FAPAR RMSE gain (%) |
|---|---|---|---|
| $GPR_{multi}$ | 2.0 | 2.1 | 2.6 |
| $NN_{multi}$ | 1.4 | 2.0 | 1.3 |
| $KRR_{multi}$ | 1.3 | 2.0 | 1.2 |

The theoretical performance of the $GPR_{multi}$ over the unseen test data is shown in the scatterplots of Figure 3 in which RMSE values of 0.68, 0.048 and 0.076 in the case of LAI, FVC and FAPAR respectively were found as well as high coefficient of determinations ($R^2 > 0.88$ in all cases). In the case of $NN_{multi}$ and $KRR_{multi}$ the results were slightly worse (not shown for the sake of brevity). In general, the scatterplots show stable estimates mainly for FVC and FAPAR. In the case of LAI, predicted values showed a saturation effect about 5 $m^2/m^2$ values, similarly to other literature works dealing with PROSAIL inversion (Bacour *et al*., 2006; Baret *et al*., 2007; Weiss *et al*., 2007; Garrigues *et al*., 2008). The residuals were relatively stable over the whole dynamic range for FVC and FAPAR whereas for LAI the regression method tends to produce an underestimation for very high LAI values (saturation domain).

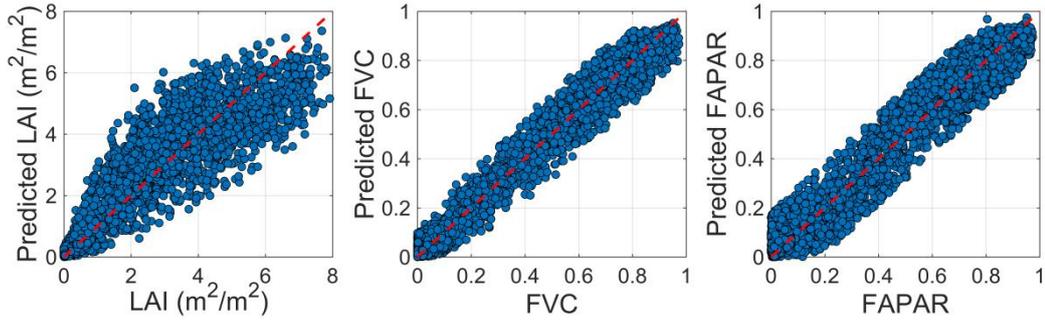

Figure 3. GPR$_{multi}$ predicted values over the unseen test set.

A sensitivity analysis was undertaken to assess the possible impact of the cardinality of the training dataset in the retrievals accuracy for all methods. We built a database composed by 7000 samples (split in 6000 for training and 1000 for testing). Then, subsets with samples of 50, 100, 200, 300, 400 500 and from 1000 to 6000 at 500 step were chosen at random in every computation for different training experiments. Figure 4 shows the RMSE variation for LAI, FVC, and FAPAR as a function of the number of training samples. It can be observed that when very few samples are used as training set (e.g., less than 100) some noise (instability) in the RMSE values is observed, which could be expected since the samples were drawn from a random distribution. As we increase the number of training samples, the models reveal better accuracies and become to stabilize. These results demonstrate that with a relatively small number of training samples (i.e. less than 2500) the RMSE becomes very stable indicating that the models did not incur in any overfitting issue and highlighting the good representativity of the simulated data.

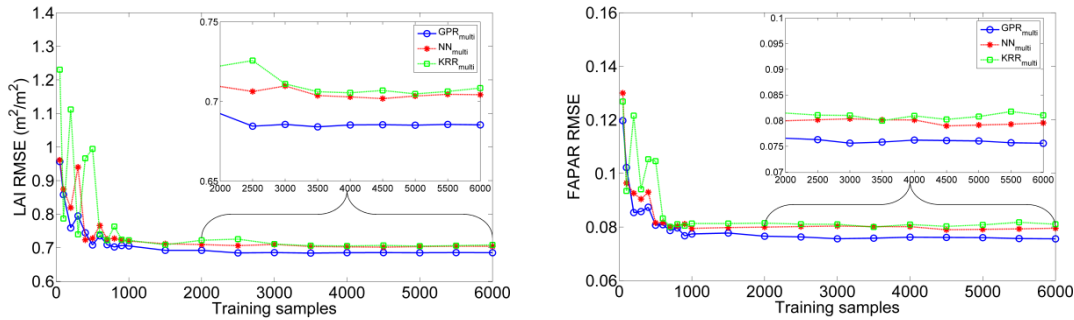

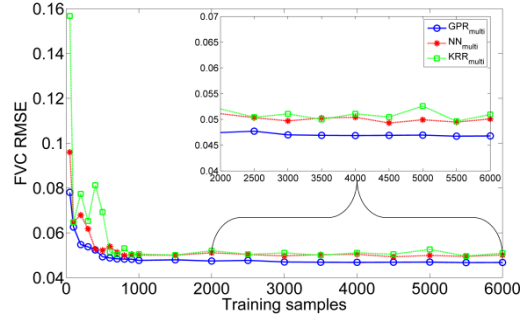

Figure 4. Variation of the RMSE for the three vegetation parameters (over the unseen test set) with the number of training samples.

The relative RMSE to range (rRMSE) was computed for every biophysical parameter (see Table 3). These results show rRMSE reductions in all cases improving the performance about 3.5% for all parameters when using the proposed GPR$_{multi}$ method.

Table 3. Relative RMSE to range $\left(rRMSE(\%) = \frac{RMSE}{(max(y_d) - min(y_d))} \times 100\right)$ obtained with the multioutput algorithms for every biophysical parameter. $y_d$ accounts for LAI, FVC and FAPAR.

| Method | LAI rRMSE (%) | FVC rRMSE (%) | FAPAR rRMSE (%) |
|---|---|---|---|
| GPR$_{multi}$ | 8.1 | 4.3 | 7.2 |
| NN$_{multi}$ | 11.6 | 7.7 | 10.7 |
| KRR$_{multi}$ | 11.8 | 7.8 | 10.8 |

In addition, we can assess the GPR$_{multi}$ gain in accuracy measuring the reduction in RMSE with regard to NN$_{multi}$ and KRR$_{multi}$. Important gains in accuracy are obtained for the joint estimation of LAI, FVC, and FAPAR. The GPR$_{multi}$ showed a gain in LAI accuracy of 4% and 5% with regard to NN$_{multi}$ and KRR$_{multi}$, respectively. In the cases of FVC and FAPAR, the GPR$_{multi}$ improvement was 4% with regard to both NN$_{multi}$ and KRR$_{multi}$.

Eventually, the multi-output models were trained 50 times for testing the model's robustness. The distribution of these results is shown in Figure 5 (left), corresponding to the histogram of LAI RMSE between estimates and test points. Figure 5 (right) shows the boxplots for the same results. These results show similar behaviour of the three methods, however, the GPR$_{multi}$ revealed as the most robust and stable regression method exhibiting smooth behaviour whereas NN$_{multi}$ and KRR$_{multi}$ presented some outliers. Similar results were obtained for FVC and FAPAR (not shown for the sake of brevity). Due to its good performance, the GPR$_{multi}$ algorithm appears to be well suited for retrieval of vegetation products and was selected in the LSA-SAF processing chain.

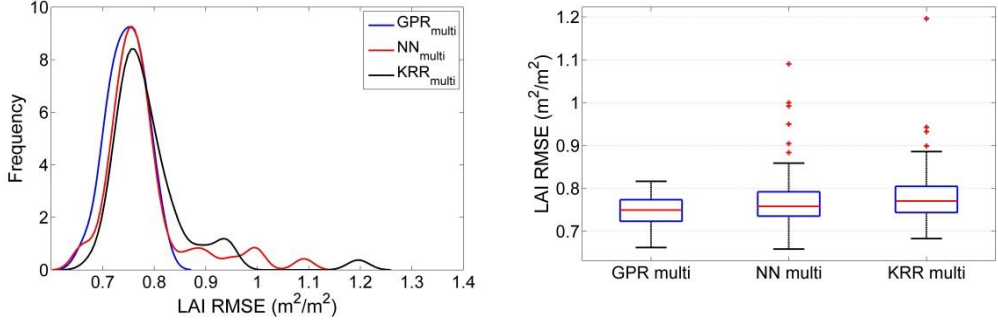

Figure 5. Distribution of RMSE (left) and associated boxplots (right) obtained running the models 50 times.

## 4 Theoretical uncertainty of retrievals

### 4.1 Interpretation of the retrievals: the vegetation isolines

The algorithm retrieves LAI, FVC and FAPAR parameters over all global land surfaces using as input ideally free of snow and ice AVHRR/MetOp BRDF ($k_0$ EPS) observations. Figure 6 shows an example of the distribution of $k_0$ EPS values in the red-NIR space over all global land surfaces masking out poor quality and ice/snow pixels. We can observe a triangle-shaped distribution of pixels called "reflectance triangle". The base of this triangle represents the "soil line" connecting pixels with little or no vegetation cover, while pixels in the opposite vertex of the triangle correspond to the densest canopies. The EPS reflectance triangle matches those provided by other satellite products, such as the MODIS BRDF model parameters (MCD43B1). Nevertheless, the $k_0$ EPS presents a slight negative bias in red and NIR channels. In fact, values out of the physical ranges are found in a few areas (i.e. less than 2% of land surface), such as in tropical forests.

Figure 6 shows the best possible prediction of FVC and LAI given a realized value of red and NIR. It results from averaging values of global maps of biophysical parameters falling in a narrow interval of red and NIR (AVHRR channels 1 and 2, respectively). Although there is not a one-to-one relationship between estimating parameter and $k_0$ field in the red-NIR domain (the algorithm uses also the middle-infrared band), the isolines allow understanding better the performance of the algorithm and interpret the pattern of estimating errors (section 4.3). The vegetation isolines capture the essential relationship between reflectance and biophysical parameters (e.g. LAI), showing a general agreement with literature studies (e.g. Huete *et al*., 1988). There is a gradual change in the value of biophysical parameters along the "reflectance triangle", a key feature to enhance the performance of the algorithm since retrieval errors are highly related to the spacing and smoothness of the isolines. The dense canopies on the top vertex present the maximum LAI values and a large dispersion (saturation domain).

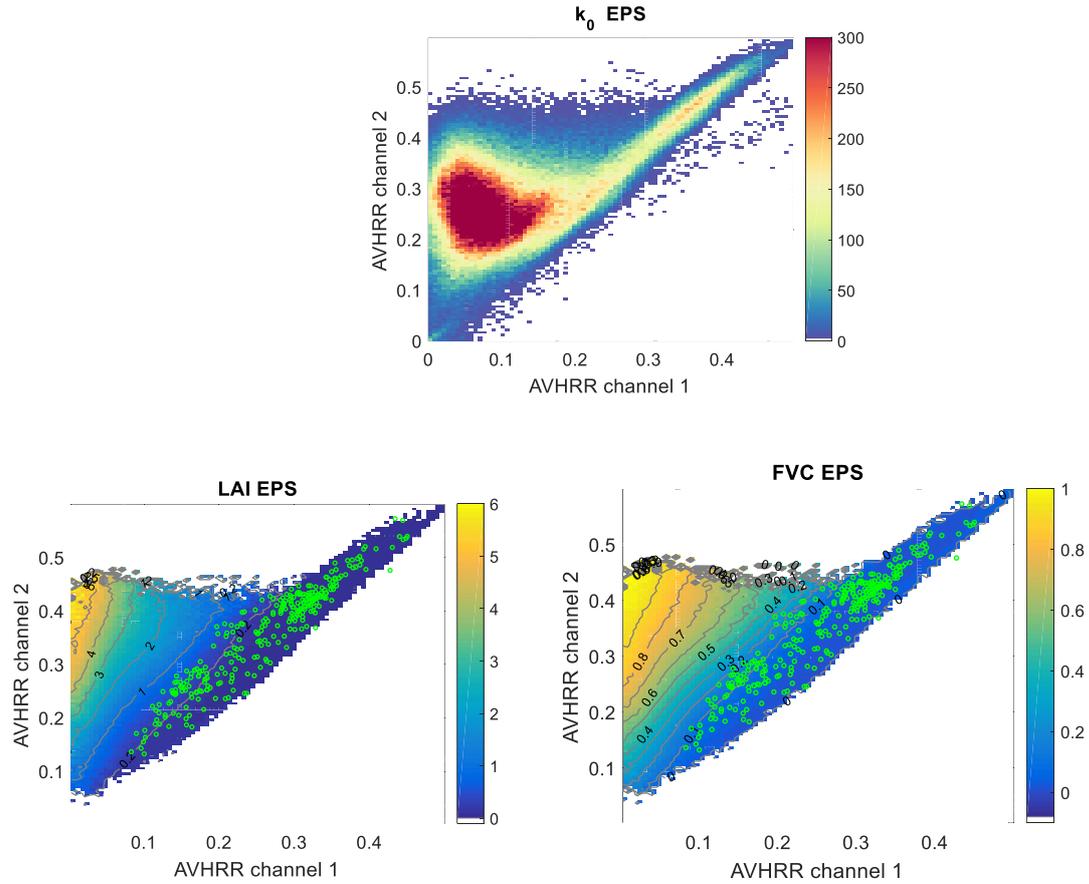

Figure 6. Top: MetOp-B spectral $k_0$ red-NIR feature space for the 15th of July 2016. Bottom: Projection of the LAI and FVC and EPS global estimates, onto the reflectance triangle. Figures were generated based on global map at a reduced (1/16 pixels) resolution. The lines of constant vegetation amounts (isolines) are drawn. Green circles correspond to the locations of the bare areas spectra used as input in PROSAIL.

## 4.2 Assessment of the product uncertainties

A quantitative uncertainty estimate is delivered with every product, namely Err(FVC), Err(LAI), Err(FAPAR). It represents the statistical confidence intervals of biophysical parameters predictions and is assessed taking into account two different sources of error, namely $\sigma_{k_0}$ and $\sigma_{GPR}$:

$$Err = \sqrt{\sigma_{GPR}^2 + \sigma_{k_0}^2} \ . \qquad (12)$$

The $\sigma_{GPR}$ is the GPR predictive standard deviation (Eq 6), which quantifies the confidence on the associated estimate. This error penalizes pixels outside of the convex hull defined by PROSAIL reflectance simulations used in the GPR model training. Hence, underrepresented spectra such outliers or dark pixels along the coastline lead to higher uncertainties for its associated retrievals.

The $\sigma_{k_0}$ error propagates the effects of the inaccuracies of the BRDF model parameters, on the prediction of biophysical parameters. Small uncertainties usually correspond to cases when a large number of observations are available during clear periods and vice verse. The error propagation was undertaken using the Monte Carlo method by computing the biophysical parameters $M$ times, each time varying the $k_0$ values randomly within their respective limits of uncertainty. $M$=100 samples were thus generated from a multivariate normal distribution drawn from the uncertainty covariance matrix of $k_0$ parameter (namely, the HDF5_LSASAF_EPS_ETAL-Channel-CK product). It should be noted that the $k_0$ covariance matrix ($C_R$) assumes that distribution of errors are Gaussian and mutually uncorrelated, i.e. $C_R$=diagonal (Err$^2$($k_0$)[C1], Err$^2$($k_0$)[C2], Err$^2$($k_0$)[C3]). Since the input $k_0$ parameters are treated as independent Gaussian variables with finite variances, uncertainty estimates by error propagation preserve the Gaussianity for Err($k_0$) errors. To speed up the computations, $\sigma_{k_0}(LAI)$, $\sigma_{k_0}(FAPAR)$ and $\sigma_{k_0}(FVC)$ were pre-computed and stored in a look-up table, using a dense distribution in the three EPS channels of $k_0$ values and their respective uncertainties, Err($k_0$).

This section is aimed to assess the main errors of estimating parameters, i.e. $\sigma_{GPR}$ and $\sigma_{k_0}$, evaluating the possible impact of the main sources of uncertainty. In particular, the influence of including uncertainties in the training data set was assessed considering varying levels of additive noise, which ranged from a null perturbation (noise free) to a strongly perturbed signal. 15 noise levels of Gaussian noise $\mathcal{N}(0, \sigma^2)$ (Eq. 1) were used with σ values equal to 0, 0.025, 0.05, 0.010, 0.015, … , 0.065.

### 4.2.1 Assessment of the $\sigma_{GPR}$ errors

The $\sigma_{GPR}$ errors were computed for all possible input values and displayed in the red-NIR domain. Results correspond to a fixed value of channel 3 (i.e. $k_0(\lambda_3)$=0.30). The $\sigma_{GPR}(LAI)$ isolines (see Figure 7) reveal that GPR prediction uncertainty for LAI is virtually constant having a value of about 0.7 for the majority of EPS pixels. These values are indicative of high confidence estimates due to the high similarity between observations and training data. We can observe as uncertainty increases abruptly ($\sigma_{GPR}$ higher than 1.0) either in the tail where bright soils situate or in very dark surfaces. In both cases, high errors are produced since $k_0(\lambda_3)$ was fixed to constant value (0.30) very different from the values found in $k_0(\lambda_3)$ for bright soils (i.e. typically higher than 0.5) and dark surfaces (i.e. typically lower than 0.15). This is an illustrative example of how uncertainty increases considerably ($\sigma_{GPR}$ above 2) in pixel reflectances that depart from training simulations. Similar results were found for FVC and FAPAR. We conclude that pixel estimates with $\sigma_{GPR}$ higher than 1.0 for LAI and 0.10 for FVC/FAPAR should be used with cautious.

Figure 7b illustrates the effect of adding noise (σ=0.015) in the training data set. A very slight increase in prediction error is found for the majority of valid observations in an EPS scene, indicating that the inclusion of moderate uncertainties in the simulations is not harmful. However, an important systematic reduction of prediction errors is found in "unrealistic" pixels: the higher the level of noise, the lower the prediction errors for surfaces not represented in the training data set. Hence adding an excessive amount of noise (e.g. σ>0.025) may be damaging due to (i) a slight loss of model specificity (i.e. a degraded capability to match signatures of valid observations) and (ii) a significant reduction of the ability to discriminate invalid observations.

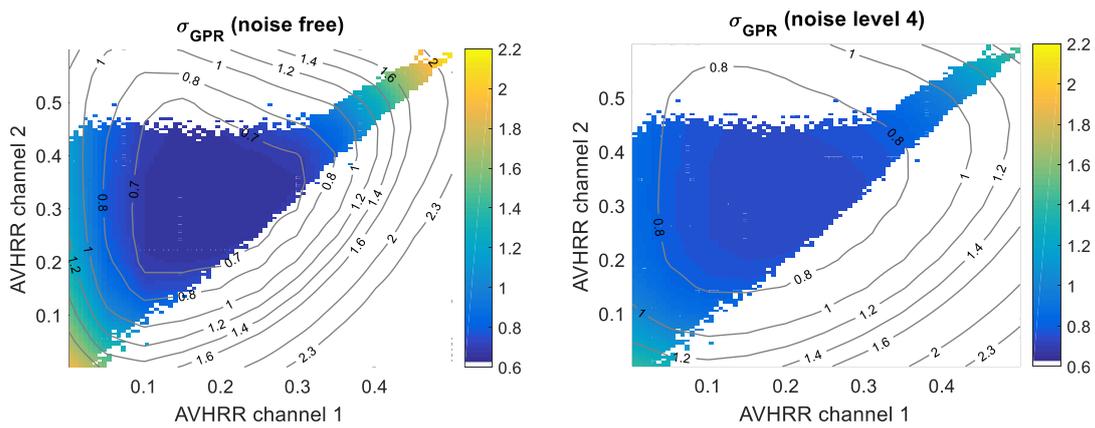

Figure 7. Prediction uncertainty for LAI algorithm ($\sigma_{GPR}$) as a function of the input normalised reflectance ($k_0$) in EPS red and NIR channels for a given value of 0.30 for MIR channel. Two different cases have been considered: noise free simulations (left panel) and simulations with a moderate level of noise, i.e. σ=0.015 (right panel). To improve the clarity, only realistic $k_0$ values (i.e. having valid observations in an EPS scene) are depicted.

### 4.2.2 Assessment of the $\sigma_{k_0}$ errors

Figure 8 shows the fields of $\sigma_{k_0}$ uncertainties, which quantify to what extent changes in the input produces changes in the estimated parameters. Results correspond to situations with a high $k_0$ error (i.e. 0.050 in all channels). Visual examination of the isolines indicates that the LAI uncertainty fields are very dependent on the pixel composition, which determines the smoothness or steepness of the LAI isolines gradient in the $k_0$ features space. The largest $\sigma_{k_0}$ errors are found in dense canopies due to saturation of reflectance, presenting a weak sensitivity to changes in vegetation properties. It is well known that saturation conditions significantly affect the reliability of derived canopy parameters, which is one major issue of retrieving LAI for dense canopies (Yan *et al*., 2016a). The contrast between vegetation and soil optical components may influence also the $\sigma_{k_0}$ error. For example, in surfaces with small red reflectances, such as dark dense canopies and wet surfaces, the LAI/FVC isolines are close together (see Figure 8), and the model is particularly sensitive to $k_0$ uncertainties leading to large LAI errors. Conversely, in bright bare areas such as deserts and semiarid regions, isolines are further apart, and the model is more robust against $k_0$ uncertainties leading to the lowest LAI errors.

Figure 8 illustrates with several examples the effects of adding different amounts of noise in the simulated reflectances. Visual examination of the results indicates that inclusion of moderate amounts of noise, such as level 2 (σ=0.005) and level 4 (σ=0.015) cause a systematic reduction of $\sigma_{k_0}$ error for all surfaces which is, however, very dependent on the surface composition. The highest improvement is observed in problematic areas with large LAI uncertainty values, such as dense canopies. The results suggest that adding noise to simulations enhances the consistency between simulations and actual satellite data, reducing overfitting issues. Conversely, the smallest $\sigma_{k_0}$ values correspond to soils, indicating that the model is very well constrained for these surfaces. Accurate retrievals are thus expected in sparse canopies. For example, zero or negligible LAI values are found in deserts and semi-desertic areas throughout the year (see maps in

Figure *11*).

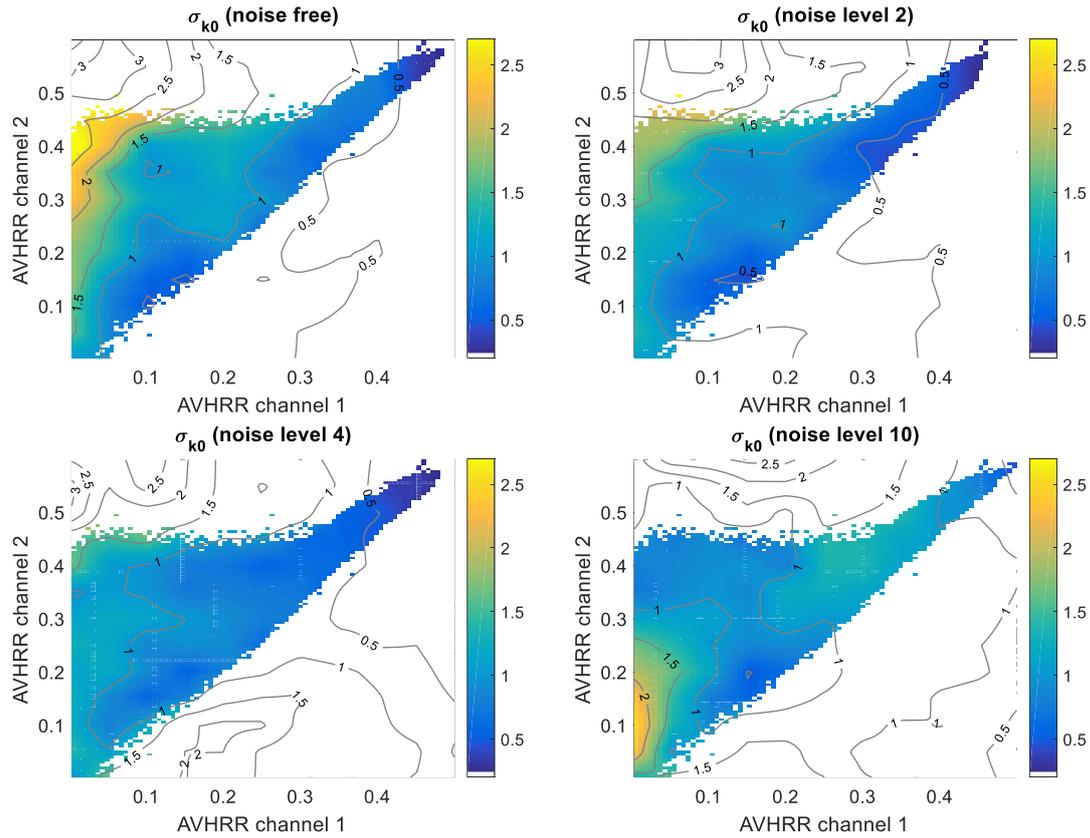

Figure 8. $\sigma_{k_0}$ (LAI) uncertainty as a function of input normalised reflectance ($k_0$) in EPS red and NIR channels for a given value of 0.20 for MIR channel. Four cases with varying noise levels in simulations are used, with σ values of 0, 0.005, 0.015 and 0.045.

In order to complete the assessment and optimise the choice of the amount of noise in the training simulations, the $\sigma_{k_0}$ (LAI) errors were computed for inputs representative of several land cover types. For each cover type, the LAI uncertainty was quantified by averaging $\sigma_{k_0}$ (LAI) for all pixels falling within the distribution of EPS $k_0$ values given by spheres with center and radius imported from a real EPS scene (see characteristics in Table 4). Several runs of the retrieval algorithm were made varying the amount of noise. The quality of the input was also examined by considering two Err($k_0$) values, 0.03 (average uncertainty) and 0.05 (i.e. poor quality).

Table 4. Characteristics of four cover types considered to compute the $\sigma_{k_0}$ in Figure 8.

| Surface | Center | Radious | Description |
| --- | --- | --- | --- |
| Dense dark vegetation | [0.03, 0.30, 0.17] | 0.02 | Closed forest ( >3m height, > 40% tree cover) |
| Dense green vegetation | [0.05, 0.42, 0.22] | 0.04 | Dense vigorous canopies (irrigated croplands, closed herbaceous cover) |

| Intermediate vegetation | [0.13, 0.35, 0.28] | 0.04 | Open shrublands, open herbaceous, mosaic cropland/grassland, savannas |
| Soil | [0.33, 0.40, 0.55] | 0.03 | Bare areas |

For the considered cases, the $\sigma_{k_0}$ (LAI) errors approximately double when Error($k_0$) goes from 0.03 to 0.05 (see Figure 9). The bare areas show the lowest errors whereas the errors in dense canopies are three times larger. Even adding very small quantities of noise, such as levels 1 ($\sigma$=0.0025) and 2 ($\sigma$=0.005) a significant retrieval error reduction is achieved. However, including high noise levels may produce degraded LAI estimates. For both dark and green dense canopies, the inclusion of moderate uncertainties in RTM simulations reduces significantly the retrieval errors, irrespectively of the quality of the input. However, the algorithm performance of both canopies is very different: for the green canopies, the retrieval error decreases monotonically with the amount of noise, while for the dark canopies a degraded performance of the algorithm is observed beyond the fifth level of noise ($\sigma$=0.020). For intermediate canopies (i.e. vegetation conditions between sparse and dense canopies), a significant gain in performance (although of lower magnitude) is found when adding slight amounts of noise but quality degrades for higher levels, particularly for average quality observations (i.e., Err($k_0$) = 0.03). Finally, for bare areas, the inclusion of noise has practically no impact on retrievals.

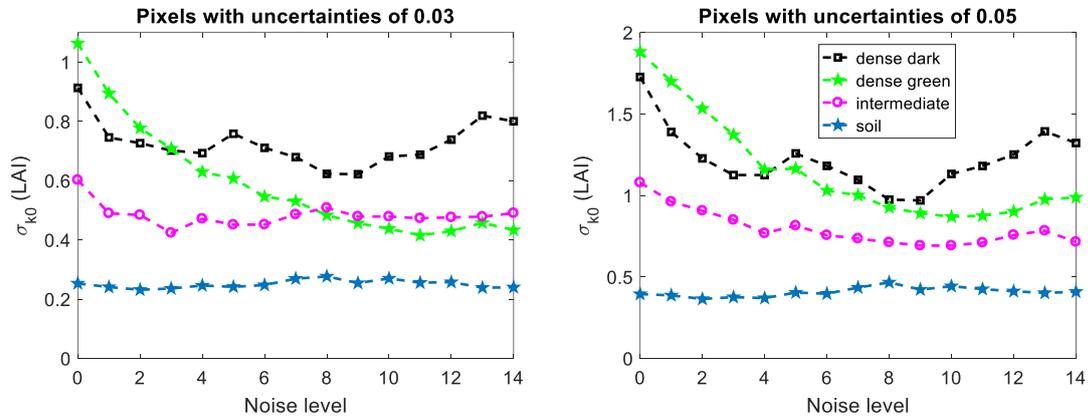

Figure 9. Assessment of the LAI error ($\sigma_{k_0}$) against the noise introduced in training database, corresponding to the four cover types described in Table 4. Two different cases of input quality are considered, with $k_0$ errors uniform in the three EPS channels: (a) Err($k_0$)=0.03; (B) Err($k_0$)=0.05

The above results reveal the convenience of adding moderate noise levels to reduce retrieval errors. By assuming some inadequacy between the PROSAIL outcomes and corresponding satellite values, simulations that are in the proximity of the actual observations are considered, which offers a means to regularize the inversion of the model and reduce overfitting. The assessment of the errors has contributed to fine tuning the algorithm by optimizing the optimal amount of noise. In particular, the fourth level of noise (σ=0.015) works reasonably well in all conditions and was chosen, with overall relative LAI error reductions of 39% for dense green canopies, 30% for dense dark canopies and 25% for intermediate canopies. Noise addition has a negligible impact on retrievals for bare and semi-arid areas, in which the most accurate and stable retrievals are obtained.

## 5  The EPS vegetation products

The multi-output GPR model, once trained for a wide range of cases for a joint optimization of the kernel hyperparameters, allows identifying the nonlinear relationship between the three AHVRR/MetOp vegetation products and atmospherically corrected cloud-cleared $k_0$ BRDF product. The AVHRR based ten-day vegetation products are generated pixel-by-pixel at a global scale, inheriting the temporal and spatial characteristics of the EPS ten-day albedo (ETAL) product, which is obtained through composite periods of 20 days (Geiger *et al.*, 2016). The LSA-SAF vegetation products are level 3 full globe rectified images in sinusoidal projection, centered at (0ºN, 0ºW), with a resolution of 1.1km×1.1km. The timeslot in the filename of this product corresponds to the last day of the 20-day time-compositing period. For example, the filename HDF5_LSASAF_M01-AVHR_ETLAI_GLOBE_201611250000 with day of production 25$^{th}$ November corresponds to the period November 6$^{th}$-25$^{th}$, 2016.

An example of the LSA-SAF EPS VEGA (FVC, LAI and FAPAR) 10-day products is shown in Figure 10. Visually, the different products are artefacts-free and spatially consistent with the available ground truth. The highest values of the different vegetation fields are systematically reached close to the Equator, such as in the Amazon Basin and Central Africa forests, followed by the Northern latitudes (e.g. around 50º in Russia) in accordance with the boreal forests distribution. The products are practically equal to 0 over the largest sand deserts, such as the Gobi, Arabian, and Sahara for which a gradual increase over the Sahel area is observed.

The FVC, LAI and FAPAR products are disseminated as a separate file coded in HDF5 format signed 16-bit integer variable, and include additional datasets and metadata attributes. The algorithm provides an estimate of the confidence or uncertainty assigned at each pixel, taking into account uncertainties due to the retrieval algorithm ($\sigma_{GPR}$) and inputs ($\sigma_{k0}$) (Eq. 12). Since the first one quantifies how close a pixel is to the training data, inspection of $\sigma_{GPR}$ maps may improve the choice of certain PROSAIL variables, mainly soil backgrounds, unaddressed in the simulation, and identify possible invalid outliers such as pixels contaminated by traces of snow/ice, undetected clouds or residual atmospheric effects. The second error ($\sigma_{k0}$) dominates in areas with large input errors where the BRDF reliability is poor, leading to unreliable estimates, mainly in dense canopies (due to saturation of reflectance). This issue is aggravated in dark surfaces since the low contrast between soil and vegetation makes the retrieval ill-conditioned. The overall error estimates, Err(FVC), Err(LAI), Err(FAPAR) (see examples in time profiles. In the medium quality areas (cyan pixels) presenting mean retrieval errors in the range 1.0-1.5 (LAI) and 0.10-0.15 (FVC/FAPAR), the observations are generally reliable but they should be used with cautious due to presence of missing data and degraded quality observations for certain specific periods. Low quality areas (orange pixels) are characterized by high uncertain retrievals, i.e. larger than 1.5 (LAI) and 0.15 (FVC/FAPAR). Finally, red colored areas corresponding to generally unusable zones are found in high latitudes due to poor illumination and atmospheric conditions, and frequent snow cover during wintertime.

Figure *11*) allow masking out problematic areas. The largest product errors correspond to unreliable inputs typically derived under sub-optimal BRDF sampling, such as in tropical forests affected by persistent cloud coverage and in high latitudes during winter due to poor illumination conditions and snow. Beyond certain uncertainty limits (e.g., 0.20 for FVC and FAPAR, and 1.5 for LAI) estimations may be regarded as unreliable and its use should be restricted.

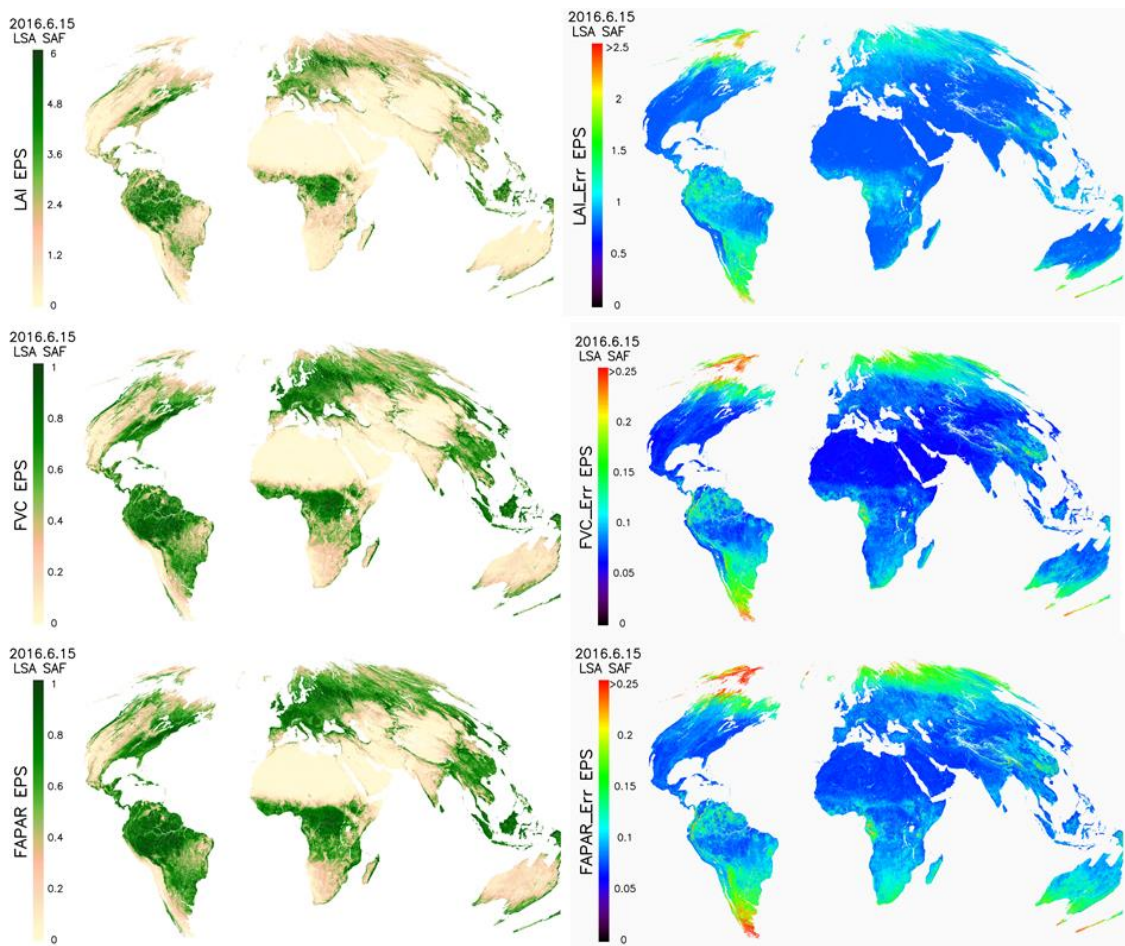

Figure 10. EPS LAI (top), FVC (middle) and FAPAR (bottom) version v1.0 products corresponding to May 26th - June 15th 2016: products (left panels) and their respective error estimates (right panels). The visualization shows the window (80°N–50°S, 100°W–140° E) covering almost all land pixels.

Figure *11*1 shows the geographical distribution of the quality for the three EPS products, as obtained from averaging their respective uncertainty estimates along the entire 2016 year. Good quality areas (green pixels) presenting mean uncertainty values below 1.0 for LAI and 0.10 for FVC/FAPAR, are consolidated regions with high spatial and temporal coverage and reliable

The seasonal variations in the quality and coverage of the products during year 2016 are depicted in

Figure *11*1. In overall, around 80% of pixels for all variables showed good and medium quality levels. Only around the 2%, 0.2% and 3.2% of Earth's land surface showed poor consistency for FVC, LAI and FAPAR, respectively. Clearly, the best performance for all products corresponds to areas with latitudes below 40ºN, such as in Africa and Australia continents. These consolidated regions generally present complete coverage of good quality observations. The temporal evolution shows the highest percentage of good quality pixels during the period from April to October, except for July and August since the LAI maximum is reached in many areas, thereby presenting higher errors (e.g. saturation domain). The largest percentage of unprocessed pixels is found during winter time in northern hemisphere, due to snow cover and/or persistent cloud coverage.

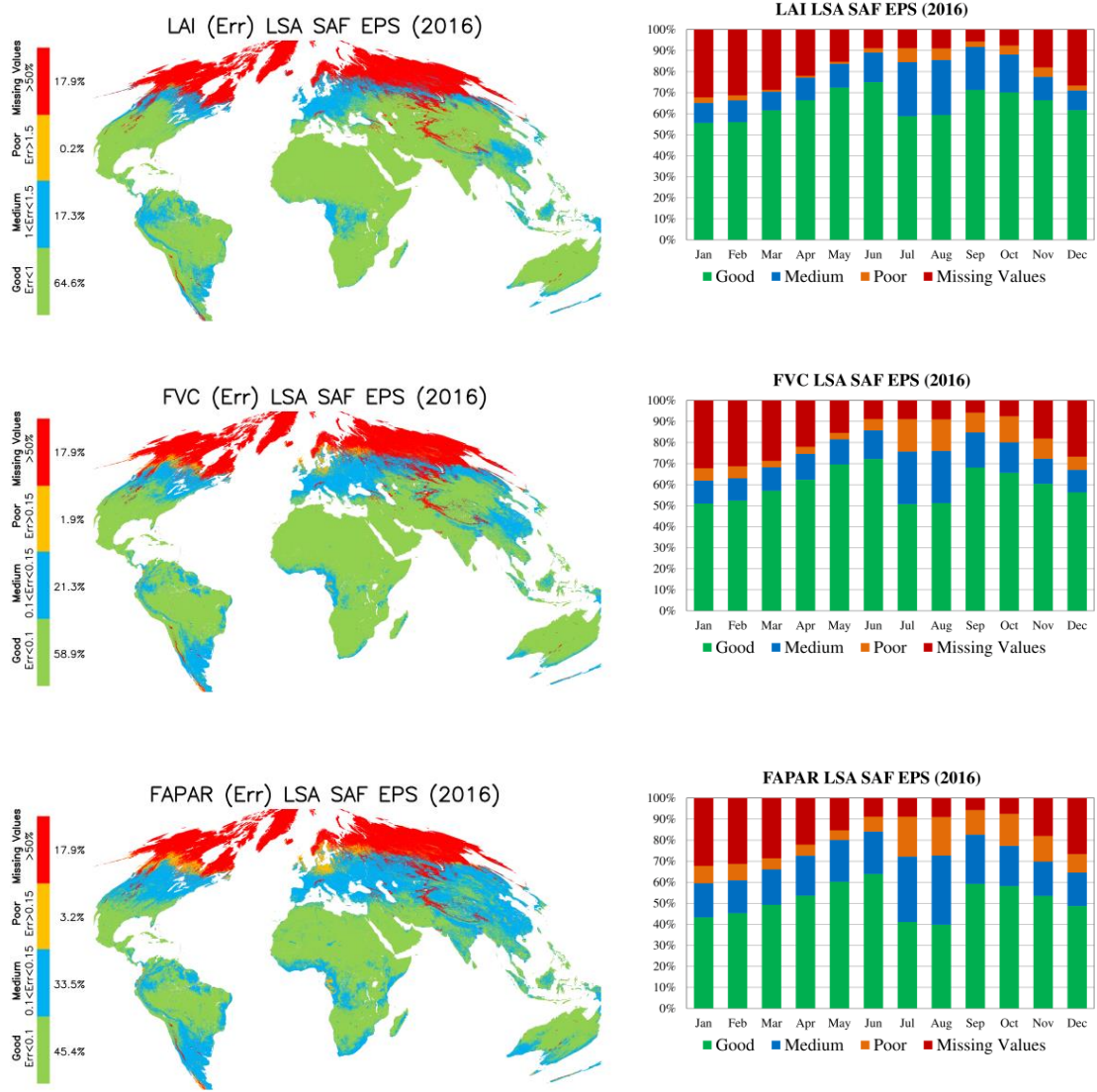

Figure 111. Left: Quality of the EPS 10-day LSA-SAF vegetation products as a function of the mean values of its theoretical uncertainty along the year 2016 (see text for details). Right: Monthly fraction of valid inland pixels for EPS 10-day vegetation products during year 2016. Percentages are classified according to three main levels of accuracy: optimal (<0.10 for FVC/FAPAR; <1.0 for LAI); medium to low ([0.10, 0.15] for FVC/FAPAR; [1.0, 1.5] for LAI); poor (>0.15 for FVC/FAPAR; >1.5 for LAI).

# 6  Validation studies

The LSA-SAF vegetation products are routinely validated. The adopted strategy for validation of EPS vegetation products (FVC, LAI and FAPAR) consists of three main steps: 1) evaluation of errors in the main variables used as input for EPS algorithm and assessment of the impact on EPS products; 2) inter-comparison with other satellite derived vegetation products; and 3) comparison with in situ measurements. This section provides a preliminary assessment of the EPS products, by examining its consistency with two reference products, MODIS Collection 6 (C6) and PROBA-V Collection 1 km Version 1 (V1).

The MODIS LAI and FAPAR (MOD15A2H) C6 products are delivered at 500 m spatial resolution and global coverage with a frequency of eight days. The retrieval algorithm chooses the best pixel available from all the acquisitions of the Terra sensor within the eight-day period. MODIS C6 uses the same retrieval algorithm as the Collection 5 at 1 km, but C6 benefited from improved surface reflectances and biome classification at an enhanced spatial resolution (Yan et al., 2016a). The uncertainties of MODIS LAI (FAPAR) C6 assessed over ground measurements are estimated to be 0.66 (0.15) (Yan et al., 2016b).

The PROBA-V V1 LAI, FAPAR and FVC products are distributed with a 10-day temporal sampling and 1/112º ground sampling distance through the Global Land Service of the European Commission's Copernicus program (http://land.copernicus.eu/global). The GEOV1 retrieval methodology relies on neural networks trained to generate the "best estimates" of LAI, FAPAR, and FVC obtained by fusing and scaling MODIS C5 and CYCLOPES (Carbon cYcle and Change in Land Observational Products from an Ensemble of Satellites) 3.1 products (Baret et al., 2013). The algorithm was developed and applied to SPOT/VGT observations until the end of the mission in May 2014, and since then it is routinely applied to PROBA-V observations. Validation results showed that SPOT/VGT V1 outperforms the quality of several similar satellite products (Camacho et al., 2013). First validation studies for PROBA-V V1 shows a good consistency with SPOT/VGT V1 but displaying a systematic overestimation of the FVC PROBA-V retrievals (Camacho et al., 2017b).

In this assessment, all datasets were geo-located to a 0.05 degree grid resolution by averaging valid observations within the native pixel values (i.e. 5×5 pixels for EPS and 12×12 pixels for MODIS C6). The PROBA-V products were resampled to the same (sinusoidal) grid to enable the comparison. A no-data value was assigned if more than 25 percent of the pixels were identified as water, snow or unreliably calculated (e.g. MODIS retrieved with backup algorithm or with significant clouds). Figure 12 shows global LAI (FAPAR) maps in the mid-June (mid-October).

For LAI, the three products present a similar spatial pattern and small differences in magnitude, mainly over areas in which the retrieval algorithms generally present good performance (bare and sparsely-vegetated areas, shrubs, herbaceous cover and cultivated areas), as it was also observed in the temporal profiles (Figure 13). The largest differences are found in the northern latitudes, since products performance is usually degraded (e.g. low illumination angles and snow contaminated pixels, see Figure 11) as well as over dense forests located on the equatorial belt (e.g. persistent cloudiness and large retrieval errors when reflectance saturates). In general, EPS shows a slight negative bias, particularly with MODIS C6 over equatorial forests, which could be partly due the lack of representation of the clumping at canopy level in the EPS algorithm. Good consistency is also achieved among the three FAPAR products (Figure 12b) and between the two (EPS and PROBA-V) FVC products (see Figure S1 in supplementary material of Appendix A).

Although the polar-orbiting satellite-based products usually present poor geographical coverage in areas with high cloud occurrence, the EPS products generally present no missing data in the tropical, subtropical and warm temperate regions, except for areas covered by snow. This is mainly because the temporal composition scheme adopted by the ETAL algorithm (Geiger et al., 2016) makes use of prior information from the previous period to avoid gaps that may be frequent in extended periods of missing data. PROBA-V and MODIS products present large regions with missing data, mostly in tropical forests (e.g. Amazonia, central Africa and Indonesia) and in the humid regions of West Africa.

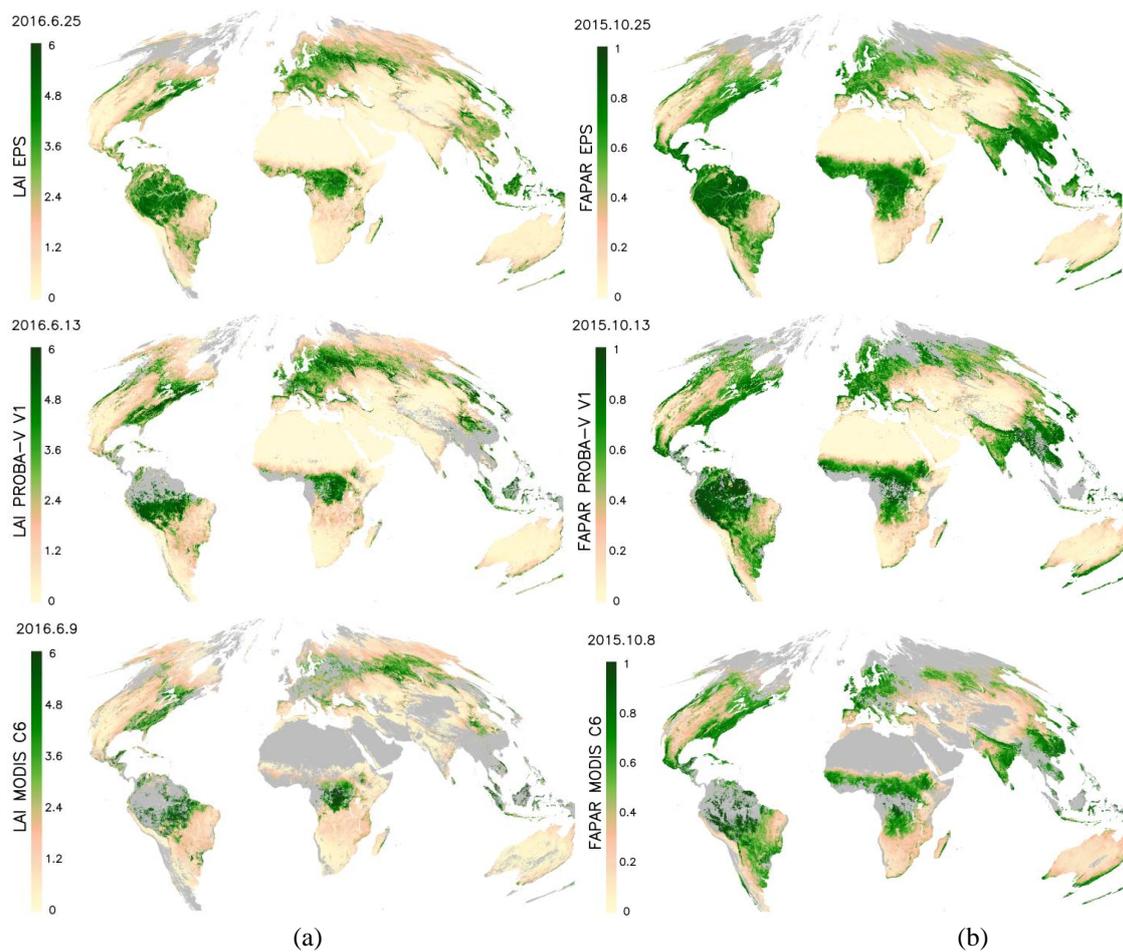

Figure 122. Spatial comparison of global LAI (a) and FAPAR (b) among EPS (top), PROBA-V (middle) and MODIS C6 (bottom) at two periods, mid-June (June 6th-25th for EPS, June 9th-16th for MODIS and May 30th-June 28th for PROBA-V) and mid-October (October 6th-25th for EPS, October 8th-15th for MODIS and September 29th - October 28th for PROBA-V). The spatial resolution is 0.05 degrees. Grey indicates missing data.

The temporal consistency of the EPS products was assessed over a network of homogeneous sites, qualitatively assessing the realism of the temporal variations of the vegetation variables and comparing them with those of equivalent MODIS (C6 and C5) and PROBA-V during the 2015-2016 period. Examples of temporal evolutions corresponding to six representative homogeneous sites are depicted in Figure 13. The intercomparison includes the previous collection (C5) of MODIS products. To assess the impact of the MODIS backup solution, the MODIS C5 temporal profiles display all valid observations whereas the MODIS C6 trajectories display only the best quality values retrieved with the main algorithm. The sites correspond to the locations of Capitanata (Italy, 41.46 ºN 15.48 ºE), Pshenichne (Ukraine, 50.07 ºN, 30.23 ºE), Homburi (Mali, 15.32 ºN, 1.54 ºW), GuyaFlux (French Guiana, 5.28 ºN, 52.91 ºW), Upper Buffalo (USA, 35.82 ºN, 93.20 ºW) and Mongu (Zambia, 15.43 ºN, 23.25 ºE). For two of the sites (Capitanata and Pshenichne) ground references were made available through the IMAGINES (Implementation of Multi-scale Agricultural Indicators Exploiting Sentinels) database (http://www.fp7-imagines.eu/). The ground measurements were collected with digital hemispherical photographs and then up-scaled up to $3 \times 3$ km$^2$ using high-spatial resolution imagery following the guidelines of the Committee on Earth Observation Satellites (CEOS) Land Product Validation (LPV) (Morisette *et al*., 2006).

The error bars associated to the EPS products are indicative of the discrepancies (systematic or random) among existing remote sensing biophysical products, and are usually larger during winter (e.g. Pshenichne site). The three products usually showed small biases and consistent inter- and intra-annual variations in the majority of biomes (crops, shrublands, herbaceous cover), with larger differences in forest areas. For Capitanata, the peak of season is reached during early spring, whereas in Pshenichne, Homburi and Upper Buffalo sites a LAI maximum is found around July. The Guyaflux (Amazonian forest) site is representative of dense tropical forests with low seasonal variations.

Although the three products are highly consistent during most of the year, EPS products present lower values Pshenichne and Upper Buffalo sites during the senescence period (September-November) under low illumination conditions leading to larger uncertainty in retrieval inputs. The two FVC products show slight differences in magnitude in the crop sites, presenting EPS lower values than PROBA-V during the peak of season. It is worth noting that EPS compares better with Pshenichne ground measurements. Both MODIS C5 and C6 show an overestimation of FAPAR in Hombury site as it was previously reported in sparse canopies (Camacho *et al*., 2013; Yan *et al*., 2016b). A significant fraction of missing data can be observed for PROBA-V and MODIS C6 products over the Guyaflux site. MODIS C5 profiles (including suboptimal retrievals with backup solution) are noisy and present unexpected temporal drops. Conversely, the EPS products present practically no missing data and thereby more continuity in the time series.

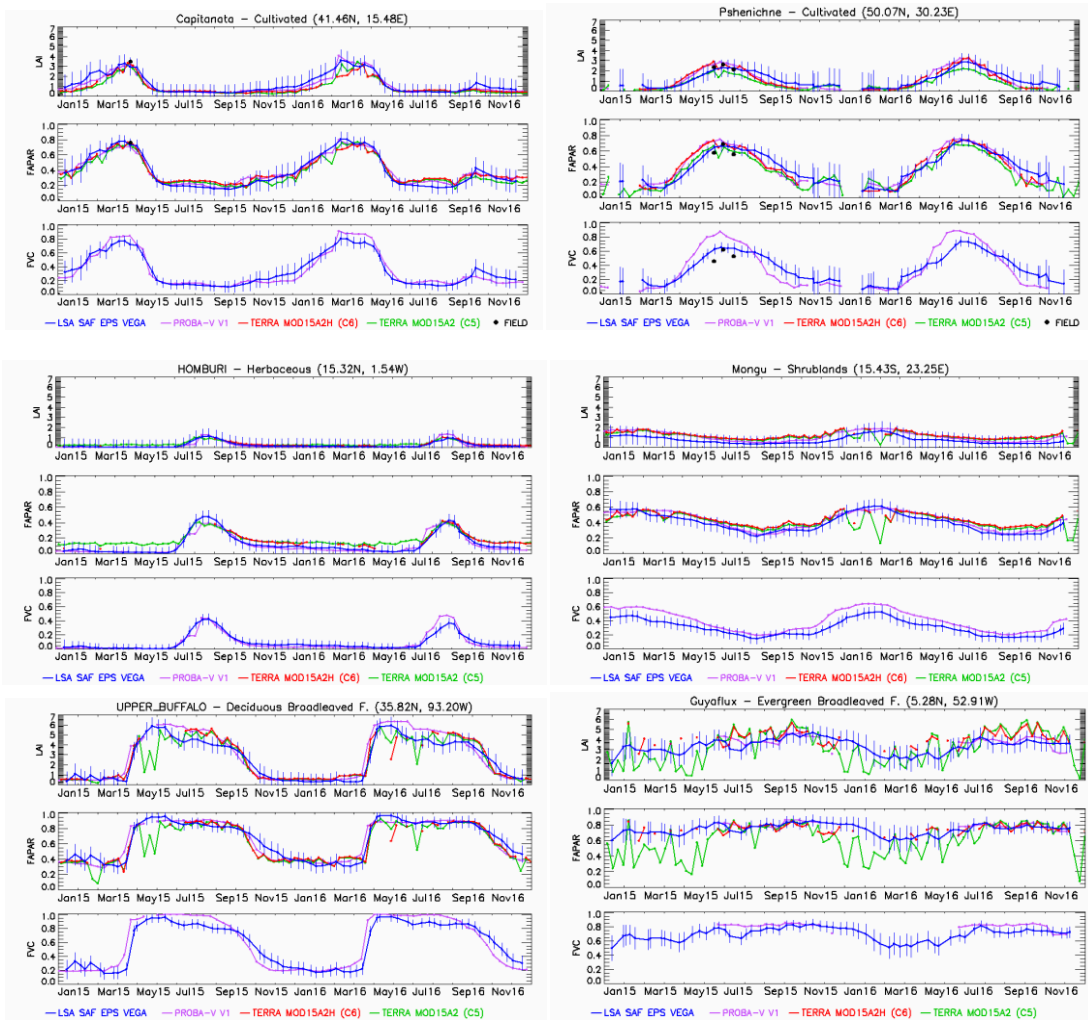

Figure 13. Temporal comparison of LAI, FVC and FAPAR among EPS, PROBA-V V1 and TERRA (MODIS C5 and MODIS C6 retrieved with the main algorithm) over representative homogeneous sites during the 2015-2016 period. Vertical bars display uncertainty associated to the EPS retrieval.

## 7 Conclusions and perspectives

A novel algorithm has been developed for the determination of global vegetation parameters based on data from the AVHRR sensor onboard MetOp satellites forming the EUMETSAT Polar System. The algorithm has been integrated into the LSA-SAF operational system and products are currently produced and delivered in near real time on a 10-day basis. After the fulfillment of an Operational Readiness Review (ORR), the LSA SAF Review Board has recommended the release of the products (LSA-403: EPS FVC, LSA-406: EPS LAI and LSA-409: EPS FAPAR) together with its user documentation to the users from the LSA-SAF website hosted at IPMA (https://landsaf.ipma.pt).

The choice of the most suitable operational EPS algorithm took into account its expected accuracy, robustness and timeliness, to solve nonlinear inversion problem at a global scale. A major advantage of the proposed multi-output approach is that it allows the joint retrieval of three parameters (LAI, FVC, FAPAR) instead of training one model per parameter using a multi-output GPR. In the multi-output inversion techniques, the biophysical parameters share the same model's parameters. Since LAI, FVC, and FAPAR are known to be highly correlated, the joint optimization of model parameters during the training may preserve relationships among the biophysical parameters in a more consistent way, increasing both prediction accuracy and computational efficiency. One further advantage is to provide a linkage between reflectances and canopy structural or leaf biochemical variables, such as dry matter content ($C_m$) and equivalent water thickness ($C_w$). Although $NN_{multi}$ and $KRR_{multi}$ proved also to be valuable algorithms, $GPR_{multi}$ outperformed them in terms of stability, accuracy, and robustness over PROSAIL EPS simulations. The GPR prediction uncertainty provides also an effective means to reject possible invalid observations, such as undetected ice/snow pixels.

The product datasets include a quantitative uncertainty estimate which is especially useful for data assimilation applications. For example, the FVC determines the partition between soil and vegetation contributions for further estimates of total emissivity and temperature, and is used as input in the operational system for generating the LSA-SAF emissivity and error budget. Performances of model inversion depend on the uncertainties associated to the satellite reflectance and the suitability of the RTM. Results have demonstrated the importance of adding a moderate amount of noise in simulations (e.g. $\sigma=0.015$) is a way of regularizing the inversion of the RTM to avoid overfitting and produce more stable solutions, reducing fluctuations caused by uncertainty in the pixel reflectance and deviations between RTM simulations and observations. Large LAI error reduction was found (i.e. >30%) in problematic areas such as dense canopies. However, adding an excessive amount of noise should be avoided since it may degrade the model performance and significantly reduce its ability to discriminate invalid observations.

A preliminary assessment of the EPS vegetation products reveals spatial and temporal consistency with equivalent (PROBA-V V1 and MODIS C6) products in most of the regions around the globe. The EPS error bars are indicative of the discrepancies among products. One main advantage for using EPS products is that they present a good spatial completeness and temporal continuity in the tropical, subtropical and warm temperate regions, mitigating known deficiencies in current operational products over cloudy areas.

LSA-SAF initiated in March 2017 its current 5-year CDOP-3 phase (Continuous Development and Operational Phase-3). During the CDOP-3 phase the EPS products will be thoroughly validated, including new validation sites and product quality verification through case studies. The feasibility to estimate a canopy water content product based on two RTM parameters, LAI and $C_w$, will be also assessed. Future work will also include the adaptation of the EPS algorithm using the MetOp-A and its reprocessing from 2008 onwards in order to obtain a long time series of homogeneous Climate Data Records. Its continuation in near real time from MetOp-C data will primarily ensure the continuity of the service during the lifetime of the European Polar System programs.

The current EPS algorithm is efficient in terms of computation time to solve nonlinear inversion problems at a global scale. Nevertheless, LSA-SAF processing system is modular and flexible, enabling the current $GPR_{multi}$ to be enhanced in future, such as designing kernel covariances that explicitly provide a linkage between outputs. During the CDOP-3 it is planned the adaptation of the current algorithm, with the development of new processing chains for EPS second generation VII and 3MI instruments. The consistent generation and distribution of EPS-based vegetation products will provide valuable and well-suited information for meteorological and environmental applications.

**Appendix A. Supplementary material**

Supplementary data associated with this article can be found, in the online version, at https://doi.org/ (to be completed by editorial staff).

**ACKNOWLEDGMENTS**

This work has been supported by the LSA SAF project, the EU under the ERC consolidator grant SEDAL-647423, and the Spanish Ministry of Economy and Competitiveness (MINECO) funding through the projects ESCENARIOS (CGL2016-75239-R) and MINECO/ERDF (TIN-2015-64210-R). Thanks are due to ImagineS project (FP7-SPACE-2012-311766) for providing ground reference data and Beatriz Fuster (EOLAB) for assisting in the preliminary validation of EPS products. The authors would like to thank their LSA-SAF colleagues from IPMA (Instituto Português do Mar e da Atmosfera) and Météo-France, the first for developing the necessary infrastructure and contributing to the integration of the vegetation algorithm into the LSA-SAF operational system and the second for developing the spectral reflectance factors used as input in the algorithm.